\documentclass[]{iopart}
\usepackage{iopams}  
\expandafter\let\csname equation*\endcsname\relax
\expandafter\let\csname endequation*\endcsname\relax
\usepackage{amsmath,amssymb,amsfonts,cite} 
\usepackage{graphicx} 
\usepackage{subfigure}
\newcommand{\be}{\begin{equation}} 
\newcommand{\ee}{\end{equation}}

\newcommand{\lp}{\left(}
\newcommand{\rp}{\right)} 
\newcommand{\bb}{\begin{bmatrix}}
\newcommand{\eb}{\end{bmatrix}}

\begin{document}

\title{The stochastic background: scaling laws and time to detection
  for pulsar timing arrays}

\author{Xavier Siemens$^{1}$, Justin Ellis$^{1}$, 
Fredrick Jenet$^{2}$, Joseph D.~Romano$^{3}$}

\address{$^{1}$Center for Gravitation, Cosmology and Astrophysics,
University of Wisconsin Milwaukee, PO Box 413, Milwaukee WI, 53201\\
$^{2}$Department of Physics and Astronomy 
and Center for Advanced Radio Astronomy,
University of Texas at Brownsville, Brownsville, TX 78520\\
$^{3}$Department of Physics and Astronomy 
and Center for Gravitational-Wave Astronomy, 
University of Texas at Brownsville, Brownsville, TX 78520}

\eads{siemens@gravity.phys.uwm.edu}

\begin{abstract} 
We derive scaling laws for the signal-to-noise ratio of the optimal
cross-correlation statistic, and show that the large
power-law increase of the signal-to-noise ratio as a function of the the observation time $T$
that is usually assumed holds only at early times. After enough time has elapsed, pulsar timing arrays enter a new regime where the
signal to noise only scales as $\sqrt{T}$. In addition, in this regime
the quality of the pulsar timing data and the cadence become relatively
un-important. This occurs because the lowest frequencies of the pulsar timing 
residuals become gravitational-wave dominated. Pulsar timing arrays enter this regime more quickly than one might
naively suspect. For $T=10$~yr observations and typical stochastic background amplitudes,
pulsars with residual RMSs of less than about $1\,\mu$s are already in that regime.
The best strategy to increase the detectability of the
background in this regime is to increase the number of pulsars in the array. We
also perform realistic simulations of the NANOGrav pulsar timing
array, which through an aggressive pulsar survey campaign adds new
millisecond pulsars regularly to its array, and show that a detection
is possible within a decade, and could occur as early as 2016.
\end{abstract}

\section{Introduction}

Pulsar timing arrays (PTAs) are sensitive to gravitational waves (GWs)
in the nano-Hz band. The most promising sources at those frequencies are super-massive binary
black holes (SMBBHs) that coalesce when galaxies merge. The
superposition of GWs from the SMBBH
mergers that have taken place throughout the history of
the universe forms a stochastic background of gravitational
waves~\cite{lb01,jb03,wl03,vhm03,ein+04,svc08,s13,McWilliams:2012jj}. These
systems may also be observable through individual periodic 
signals~\cite{svv09,sv10,rs11,rwh+12,Mingarelli:2012hh}
and bursts~\cite{vl10,Cordes:2012zz}. A number of data analysis techniques have
been implemented to search pulsar timing data for the stochastic
background~\cite{det79,srt+90,l02,jhl+05,jhs+06,abc+09,hlm+09,%
ych+11,vhj+11,Cordes:2011vg,dfg+12,vh13,Lentati:2012xb,Taylor:2012vx,esvh13},
periodic signals~\cite{jll+04,yhj+10,cc10,lwk+11,%
ejm12,bs12,esc12,pbs+12}, and bursts~\cite{fl10}.

Although the stochastic background produces random changes in the
times-of-arrival (TOAs) of pulses from individual pulsars, 
its effects on the cross-correlation of timing residuals of two pulsars only depends on the angular separation
between them~\cite{hd83}. In this paper we study the scaling
properties of the optimal cross-correlation statistic 
introduced in~\cite{abc+09}. This cross-correlation is optimized for
the spectrum of the GW background, and accounts for the intrinsic
noise of the pulsars. Using this statistic we also calculate the detection
probability for the NANOGrav pulsar timing array,
finding that a detection could occur as early as 2016, and is likely
before the end of the decade.

This paper is organized as follows. In Section~\ref{sec:signal} we
review the cross-correlation statistic and its signal-to-noise
ratio (SNR). In Section~\ref{sec:crosscorr} we derive analytic scaling laws
for the SNR of a simplified PTA as a function of the various properties and
characteristics of the array. In Section~\ref{sec:Timetodetect}
we perform realistic numerical simulations, including red spin noise, 
of the NANOGrav pulsar timing array. 
We conclude in Section~\ref{sec:conc}.

\section{The optimal cross-correlation statistic}
\label{sec:signal}

In~\cite{esvh13} we showed that for a timing array consisting of $M$
pulsars, a simple linear  (though non-invertible) transformation
allows us to write the likelihood function for the
residuals as a multivariate Gaussian
\be 
\label{eq:liker}
p(\mathbf{r}|\vec\theta)=\frac{1}{\sqrt{\det(2 \pi
\boldsymbol{\Sigma})}}\exp\lp-\frac{1}{2}\mathbf{r}^T\boldsymbol{\Sigma}^{-1}\mathbf{r}\rp,
\ee 
where 
\be 
\mathbf{r}=\bb {r}_{1} \\ {r}_{2}\\ \vdots \\{r}_{M} \eb 
\ee 
is a vector of the timing residuals time-series, $r_{I}(t)$, for all
pulsars,  $\vec\theta$ is a vector of noise and signal parameters,
\be 
\boldsymbol{\Sigma} = \langle \mathbf{r}
\mathbf{r}^T \rangle = \mathbf{R}\langle \mathbf{n} \mathbf{n}^T
\rangle \mathbf{R}^T= \mathbf{R}\boldsymbol{\Sigma}_{n}\mathbf{R}^{T}
\ee 
is the covariance matrix of the residuals, $\boldsymbol{\Sigma}_{n}$
is the covariance matrix for the pre-fit underlying Gaussian process
$\mathbf{n} $ which contains the gravitational waves along with other sources
of noise, and 
\be 
\label{eq:rmat}
\mathbf{R}=\bb R_{1} & 0 & \hdots & 0\\ 0 & R_{2} & \hdots & 0\\
\vdots & \vdots & \ddots & \vdots\\ 0 & 0 & \hdots & R_{M}\eb. 
\ee
is the timing model fitting matrix~\cite{d07}.
The covariance matrix for the timing residuals is the block matrix
\be 
\label{eq:cov} \boldsymbol{\Sigma}=\bb P_{1} & S_{12} & \hdots &
S_{1M}\\ S_{21} & P_{2} & \hdots & S_{2M}\\ \vdots & \vdots & \ddots &
\vdots\\ S_{M1} & S_{M2} & \hdots & P_{M}\eb, 
\ee 
where 
\begin{align}
P_{I}&=\langle r_{I}r_{I}^{T}\rangle,\\
S_{IJ}&=\langle
r_{I}r_{J}^{T}\rangle\big|_{I\ne J}
\end{align} 
are the auto-covariance and cross-covariance matrices for
each set of residuals. These matrices can be constructed by taking
inverse Fourier transforms of the frequency domain auto- and
cross-power spectra, and acting on them with the $R$-matrices,
\be 
\langle r_I r_I^T \rangle_{ij} = R_I \left[\int_{-\infty} ^{\infty} df e^{2\pi i f (t_i-t_j)} P_I(f) \right]R^T_I
\label{auto}
\ee
and
\be 
\langle r_I r_J^T \rangle_{ij} = \chi_{IJ} R_I \left[\int_{-\infty} ^{\infty} df e^{2\pi i f (t_i-t_j)} P_g(f) \right]R^T_J
\label{cross}
\ee
where $ij$ are matrix component indices, $P_I(f)$ is the power
spectrum of the $I$th pulsar, $P_g(f) $ is the GW power spectrum, and $\chi_{IJ}$ are the Hellings-Downs coefficients.
This treatment is somewhat different
from previous Bayesian analyses of the likelihood~\cite{hlm+09,vl10,vhj+11}. Our
treatment amounts to a \emph{conditional} pdf whereas~\cite{hlm+09,vl10,vhj+11} used a \emph{marginalized}
pdf. See~\cite{esvh13} for more details.

In~\cite{abc+09}, we showed that in the weak-signal limit the 
likelihood ratio maximized over
amplitudes leads to the optimal cross-correlation statistic for a
PTA. The optimal statistic is
\be
\hat A^2 =\frac{\sum_{IJ}r^T_I P_I^{-1}\tilde S_{IJ}P_J^{-1} r_J}
{ \sum_{IJ} {\rm Tr} \left[ P_I^{-1}\tilde S_{IJ}P_J^{-1}\tilde S_{JI} \right]},
\ee
where $\sum_{IJ}$ is the sum over all pulsar pairs. The normalization factor  
$\left(\sum_{IJ}{\rm Tr}\left[P_I^{-1}\tilde S_{IJ}P_J^{-1}S_{JI} \right]\right)^{-1}$
is chosen so that $\hat A^2$ is also the maximum likelihood estimator for the amplitude of the
stochastic background $A^2$, and the amplitude independent cross-correlation matrix $\tilde S_{IJ}$ is defined by
\be
A^2 \tilde S_{IJ}=S_{IJ}=\langle r_I r_J^T \rangle.
\ee

In the absence of a cross-correlated signal (or if the signal is weak)
the expectation value of $\hat A^2$ vanishes and its
standard deviation is~\cite{abc+09}
\be
\sigma_0 =\left(\sum_{IJ} {\rm Tr} \left[ P_I^{-1}\tilde S_{IJ}P_J^{-1}\tilde S_{JI} \right]\right)^{-1/2},
\ee
so if in a particular realization we measure a value $\hat A^2$ of
the optimal statistic, the SNR for that realization is
\be
\rho=\frac{\hat A^2}{\sigma_0}=\frac{\sum_{IJ}r^T_I P_I^{-1}\tilde S_{IJ}P_J^{-1} r_J}
{ \left(\sum_{IJ} {\rm Tr} \left[ P_I^{-1}\tilde S_{IJ}P_J^{-1}\tilde S_{JI}
  \right] \right)^{1/2} }.
\label{eq:SNR}
\ee
This definition of the SNR measures our degree of surprise (in
standard deviations away from 0),  of finding
cross-correlations in our data.
The average SNR is
\be
\langle \rho \rangle = A^2  \left( \sum_{IJ} {\rm Tr}
  \left[ P_I^{-1}\tilde S_{IJ}P_J^{-1}\tilde S_{JI} \right] \right)^{1/2}.
\label{rhoav}
\ee

It is worth pointing out that this SNR only involves
the cross-correlation terms of the likelihood, and does not include
the auto-correlation terms.  It is therefore quite a stringent test of
the presence of the stochastic background compared to, say, a
likelihood based analysis that contains contributions to the likelihood from both auto- and
cross-correlations~\cite{vlm+09,Lentati:2012xb,Taylor:2012vx,esvh13}. There
is a very good reason for doing this: red noise in the auto-correlations
could arise from intrinsic spin noise, frequency noise, or from
propagation in the interstellar medium. The only way to be certain
that we have detected a stochastic background of gravitational
waves is to ensure the data are cross-correlated as we expect. From a
Bayesian perspective we could achieve the same thing by comparing the evidence of a
model with GW red noise (with correlated and uncorrelated components),
with that of a model with only uncorrelated red noise. We can think of the SNR of the optimal
cross-correlation statistic as a proxy for such an analysis.

\section{Scaling laws for the cross-correlation statistic}
\label{sec:crosscorr}

Equation~\ref{rhoav} for the SNR can be evaluated in
the time domain for any PTA
configuration (cadence, intrinsic red and white noises, etc.), along with a GW background
amplitude and slope. The auto-covariance and cross-covariance matrices can
be constructed by taking inverse Fourier transforms of the power
spectra and cross-spectra, respectively, and accounting for fitting of
the timing model in the time-domain as shown in Eqs.~\ref{auto} and \ref{cross}. 

To understand how the SNR scales with the various properties of the PTA, however, it
is easier to work in the frequency domain. The timing model,
involves, among other things, subtracting out a quadratic (by fitting
the period and period derivative out of the TOAs). In the frequency
domain this can be crudely approximated by a low
frequency  cutoff at $f_L=1/T$, where $T$ is the time-span of the
data. We can express the trace in the sum of Eq.~\ref{rhoav} in the
frequency domain as follows~\cite{abc+09}
\be
{\rm Tr}\left[P_I^{-1}\tilde
  S_{IJ}P_J^{-1}\tilde S_{JI} \right] 
=\frac{2 \chi_{IJ}^2 T}{A^4} \int_{f_L}^{f_H} df\, \frac{P_g^2(f)}{P_I(f) P_J(f)}
\ee
where $\chi_{IJ}$ are the Hellings Downs coefficients and the
gravitational wave spectrum is 
\be
P_g(f) = \frac{A^2}{24 \pi^2} \left(\frac{f}{f_{\text{ref}}} \right)^{2\alpha}f^{-3} = bf^{-\beta},
\ee
where $f_{\rm ref}$ is some reference frequency which we will take
here to be 1~yr. This means we can write the average SNR for the PTA as 
\be 
\langle \rho \rangle = \left(2 T \sum_{IJ} \chi_{IJ}^2
  \int_{f_L}^{f_H} df\, \frac{P_g^2(f)}{P_I(f) P_J(f)} \right)^{1/2}.
\label{intSNR}
\ee
If we assume there is only white noise in our pulsars (in addition to
GWs), and that the data
are evenly sampled with cadence $c=1/\Delta t$, 
namely $P_I(f)=P_g(f)+2 \sigma_I^2 \Delta t$,
the integral in Eq.~\ref{intSNR} can be written in terms of hypergeometric functions as follows 
\be
\begin{split}
\int_{f_L}^{f_H} df\, \frac{b^2
  f^{-2\beta}}{\left(bf^{-\beta}+2\sigma_I^2 \Delta t \right)
  \left(bf^{-\beta}+2\sigma_J^2 \Delta t \right)}
=\frac{1}{\sigma_I^2-\sigma_J^2}  
\\
\times \Bigg[ f_H \sigma_I^2 G \left(-\frac{2  \Delta t f_H^\beta \sigma_I^2}{b}\right) -
f_L \sigma_I^2 G \left(-\frac{2  \Delta t f_L^\beta \sigma_I^2}{b}\right) 
\\
-f_H \sigma_J^2 G \left(-\frac{2 \Delta t f_H^\beta \sigma_J^2}{b}\right)+
f_L \sigma_J^2  G \left(-\frac{2\Delta t f_L^\beta \sigma_J^2}{b}\right) \Bigg],
\end{split}
\ee
where $G(x)={}_2F_1 (1,\beta^{-1}, 1+\beta^{-1}, x)$, and $_2F_1$ is the
hypergeometric function. We will investigate the use of these solutions in future work. 

For now let's consider the more simple situation where
all the pulsars in our timing array have the same
intrinsic white noise with RMS $\sigma$, and that the data
are evenly sampled with cadence $c=1/\Delta t$, i.e. we let
$P_I(f)=P_J(f)=P(f)=P_g(f)+2 \sigma^2 \Delta t$. This means the
average SNR can be written as the product
\be 
\langle \rho \rangle = \left(\sum_{IJ} \chi_{IJ}^2 \right)^{1/2}
\left(2T \int_{f_L}^{f_H} df\,\frac{b^2
    f^{-2\beta}}{\left(bf^{-\beta}+2\sigma^2 \Delta t \right)^2} \right)^{1/2}.
\label{eq:SNRallsame}
\ee
It follows from this expression that the average SNR scales like the
number of pulsars $M$ in the timing array, since 
$\langle \rho \rangle \propto \left( \sum_{IJ} \chi_{IJ}^2 \right)^{1/2} \propto M$.
The integral we need to evaluate in Eq.~\ref{eq:SNRallsame} is
\be 
\int_{f_L}^{f_H} df\, \frac{b^2
  f^{-2\beta}}{\left(bf^{-\beta}+2\sigma^2 \Delta t \right)^2},
\label{eq:sigint}
\ee
which we show below can be expressed in terms of hypergeometric
functions.

Before we do this, it is useful to consider two regimes in which Eq.~\ref{eq:sigint} can be easily
evaluated. The first is the {\it weak-signal limit} where the intrinsic white noise
dominates $2 \sigma^2\Delta t \gg bf^{-\beta}$ for all frequencies $f \in
[f_L,f_H]$. In this case Eq.~\ref{eq:sigint} becomes, 
\be 
\frac{b^2}{(2  \sigma^2 \Delta t)^2}\int_{f_L}^{f_H} df\,  f^{-2\beta}
\approx  \frac{b^2}{(2   \sigma^2 \Delta t)^2} \frac{T^{2 \beta -1}}{2\beta-1},
\label{eq:sigint2}
\ee
because $f_L=1/T$, so that the SNR becomes
\be 
\langle \rho \rangle = \left(\sum_{IJ} \chi_{IJ}^2 \right)^{1/2}
\frac{b c T^\beta}{\sigma^2 \sqrt{(4\beta-2)}}.
\label{eq:weaksignal}
\ee
The opposite regime,  the {\it strong-signal limit}, occurs when the GW power is
larger than the white noise, $2 \sigma^2\Delta t \ll bf^{-\beta}$,
for all frequencies $f \in [f_L,f_H]$. In this case
Eq.~\ref{eq:sigint} simply becomes, 
\be 
\int_{f_L}^{f_H} df \approx f_H = \frac{1}{2 \Delta t}=\frac{c}{2},
\label{eq:sigint5}
\ee
where we've set the high frequency cutoff $f_H$ to be the Nyquist frequency. So the
SNR becomes
\be 
\langle \rho \rangle = \left(\sum_{IJ} \chi_{IJ}^2 \right)^{1/2}
\left( c T\right)^{1/2}.
\label{eq:strongsignal}
\ee

\begin{figure}[ht]
  \begin{center}
	\includegraphics[scale=0.3]{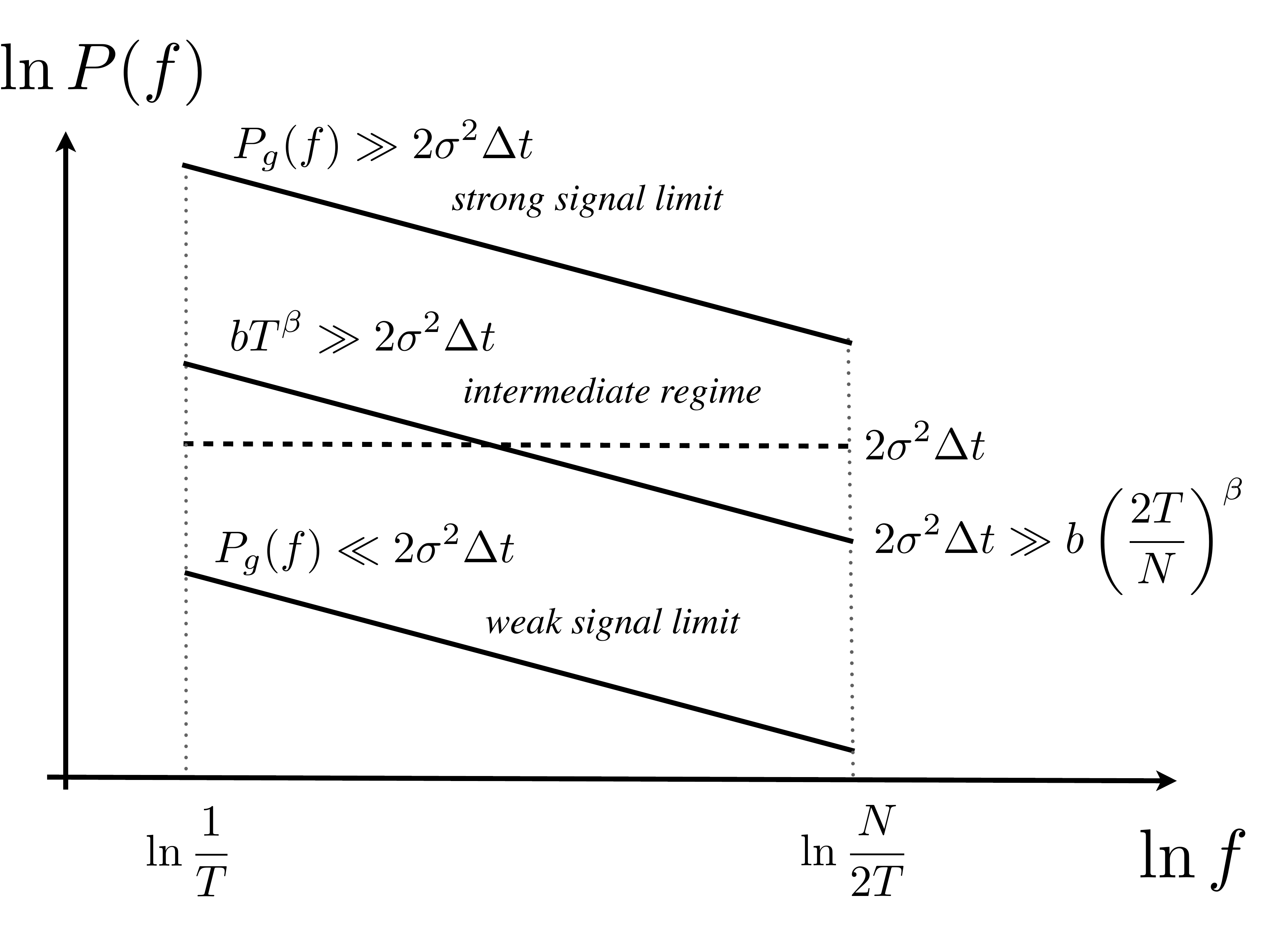}
  \caption{Schematic of the three scaling regimes. The weak-signal limit, 
    where Eq.~\ref{eq:weaksignal} is valid, is appropriate when the power spectrum of the
    stochastic background is lower than the white noise for all
    frequencies. The strong-signal limit, where
    Eq.~\ref{eq:strongsignal} is valid, is appropriate in the opposite
    limit, when the power spectrum of the background is larger than
    the white noise at all frequencies.  The intermediate regime,
    where Eq.~\ref{eq:intermediate} is valid, is appropriate when the
    lowest frequencies of the background are above the white noise
    level.  
  }
 \end{center}
\label{fig:threeregs}
\end{figure}

There is an {\it intermediate regime}, where the low frequencies of
the background are above the white noise level, but the highest
frequencies are below it. In this regime $P_g(f_L)=b T^\beta \gg 2 \sigma^2 \Delta
t$ and $2 \sigma^2 \Delta t \gg b (2T/N)^\beta =P_g(f_H)$. The integral
involved in the SNR calculation, Eq.~\ref{eq:sigint}, can be written as 
\be 
\int_{f_L}^{f_H} df\, \frac{b^2
  f^{-2\beta}}{\left(bf^{-\beta}+2\sigma^2 \Delta t \right)^2}=\int_{0}^{f_H} df\, \frac{b^2
  f^{-2\beta}}{\left(bf^{-\beta}+2\sigma^2 \Delta t \right)^2}-\int_{0}^{f_L} df\, \frac{b^2
  f^{-2\beta}}{\left(bf^{-\beta}+2\sigma^2 \Delta t \right)^2},
\label{eq:sigint6}
\ee
and each of these integrals evaluated via
\be
\begin{split}
\int_{0}^{f_*} df\, \frac{b^2
  f^{-2\beta}}{\left(bf^{-\beta}+2\sigma^2 \Delta t \right)^2}&=\\
&\frac{f_*}{\beta}\left(\frac{1}{1+
\frac{2 \sigma^2 \Delta t f_*^\beta}{b}}+(\beta-1) {}_2F_1(1,\beta^{-1}, 1+\beta^{-1},
  -\frac{2 \sigma^2 \Delta t f_*^\beta}{b}\right).
\end{split}
\ee
The argument of the hypergeometric function for the second integral of
the RHS of Eq.~\ref{eq:sigint6} is small, and the hypergeometric function can be
approximated to be unity. For the first integral, the situation is
somewhat more complicated, the argument is large and we
need to make an asymptotic expansion of the hypergeometric
function. We give details of this calculation
elsewhere~\cite{ccde+13}, but the result is
\be 
\langle \rho \rangle = \left(\sum_{IJ} \chi_{IJ}^2 \right)^{1/2}
\left[ 2 \left(\alpha \left(\frac{bc}{2\sigma^2}\right)^{1/\beta}T-1
  \right) \right]^{1/2},
\label{eq:intermediate}
\ee
where
\be
\alpha=\frac{\beta-1}{\beta}\Gamma(1-\beta^{-1}) \Gamma(1+\beta^{-1})\,.
\label{eq:alpha}
\ee

Figure~1 shows a plot of the gravitational wave
power spectrum for all three regimes as well as the white noise. 
The transition from the weak signal regime to the intermediate regime 
occurs when the lowest frequency
bin of the background becomes larger than the level of the white
noise, i.e. when $bT^\beta > 2 \Delta t \sigma^2$, the condition on
the RMS is then
\be
\sigma < \frac{A}{\pi}\sqrt{\frac{cT^\beta}{48}}.
\label{RMSthreshold}
\ee
For typical pulsar timing experiment durations of $T=5$~yr, and
cadence of about $c=20$~yr$^{-1}$, for a background with amplitude
$A=10^{-15}$ and a spectral index like the one we expect for the SMBBH
background ($\beta=13/3$), the pulsar timing array is in the weak signal limit only
if the pulsars have white noise RMSs greater than about $215$~ns. Note
that of the 17 pulsars analyzed in the 5~yr data set analyzed
in~\cite{dfg+12}, 9 have residuals with RMSs smaller than 215~ns. 

Pulsar timing arrays currently have at least a few pulsars with RMS residuals
considerably smaller than that, and Eq.~\ref{eq:weaksignal} cannot be
used. Table~\ref{table:RMSsAandT} shows the white noise RMS thresholds
between the weak and intermediate regimes for low-, mid-, and upper
ranges of the amplitude of the background and observation times $T$ of 5,
10, 15, and 20 years.

It is worth pointing out that even though the lowest frequencies of
the GW background may be larger than the white noise, the
red noise induced residuals need not be large. The red noise induced
residuals are
\be
\sigma^2_{\rm red}=\int_{f_L}^{f_H} df\, bf^{-\beta} \approx \frac{bT^{\beta-1}}{\beta-1},
\ee
for $\beta>1$, so
\be
\sigma_{\rm red} \approx\frac{A}{\pi} \sqrt{\frac{T^{\beta-1}}{24(\beta-1)}}.
\ee
For $A=10^{-15}$, $\beta=13/3$, and $T=5$~yr, the red noise induced
RMS is $\sigma_{\rm red} \approx 17$~ns, which is considerably smaller
than the 215~ns RMS of {\it white} noise below which we enter the
intermediate regime (see Table \ref{table:RMSsAandT}).  Note
that the pulsar with the smallest residuals analyzed 
in~\cite{dfg+12}, J1713+0747, had a timing residual RMS of 30~ns.

\begin{table}
\begin{center}
\begin{tabular}{c | c c c c}
 & 5 \text{ yr} & 10 \text{ yr} & 15 \text{ yr} & 20 \text{ yr}\\
\hline 
$5.6 \times 10^{-16}$ & $120$ ns & $540$ ns & $1.3$ $\mu$s & $2.4$ $\mu$s \\
$1 \times 10^{-15}$ & $215$ ns & $965$ ns & $2.3$ $\mu$s & $4.3$ $\mu$s\\
$2 \times 10^{-15}$ & $430$ ns & $1.9$ $\mu$s& $4.6$ $\mu$s & $8.7$ $\mu$s\\
\end{tabular}
\caption{White noise RMS values (using Eq.~\ref{RMSthreshold}) at which we
transition from the weak-signal limit to the intermediate regime for
low, mid-range, and upper values of the SMBBH background~\cite{s13}, $\beta=13/3$, and
four observation times. We have taken the cadence to be
$c=20$~yr$^{-1}$.  PTA data time-spans are now reaching 20~yrs, so
pulsars with white noise levels of less than a few $\mu$s are already
in the intermediate regime.}
\label{table:RMSsAandT}
\end{center}
\end{table}

In practice the strong-signal regime is irrelevant. We enter this regime
when the highest frequency bin of the background is larger than the
white noise,  i.e. that $b(2T/N)^\beta > 2 \Delta t \sigma^2$. In
terms of the RMS this condition is 
\be
\sigma < \frac{A}{\pi}\sqrt{\frac{1}{24}}\left(\frac{2}{c}\right)^{(\beta-1)/2}.
\ee
Using a cadence of $c=20$~yr$^{-1}$, for a background with amplitude
$A=10^{-15}$ and a spectral index like the one we expect for the SMBBH
background, $\beta=13/3$, the condition on the RMS is $\sigma < 4.3
\times 10^{-11}$~s. The constraint can be loosened only by reducing the
cadence, which reduces the highest frequencies probed by the experiment.

To summarize, the SNR scales with PTA characteristics in two different ways
depending on whether the lowest frequencies of the stochastic
background power spectrum are above or below the white noise level. 
When $P_g(f) \ll 2 \sigma^2\Delta t $ for all frequencies $f
\in[f_L,f_H]$ the SNR is given by 
\be 
\langle \rho \rangle = \left(\sum_{IJ} \chi_{IJ}^2 \right)^{1/2}
\frac{b c T^\beta}{\sigma^2 \sqrt{(4\beta-2)}} \propto Mc\frac{A^2}{\sigma^2}T^\beta.
\label{eq:weaksignal2}
\ee
If the PTA begins in this regime, after enough time passes, the condition $bT^\beta < 2 \Delta t
\sigma^2$ is no longer satisfied and the lowest frequency bins of the
background become larger than the white noise. At this stage the SNR scaling changes to
\be 
\langle \rho \rangle = \left(\sum_{IJ} \chi_{IJ}^2 \right)^{1/2}
\left[ 2 \left(\alpha \left(\frac{bc}{2\sigma^2}\right)^{1/\beta}T-1
  \right) \right]^{1/2} \propto M
\left(\frac{A}{\sigma \sqrt{c}}\right)^{1/\beta} T^{1/2}\,,
\label{eq:intermediate2}
\ee
with $\alpha$ given by Eq.~\ref{eq:alpha}. Table~\ref{table:RMSsAandT} shows the white noise RMS
thresholds below which Eq.~\ref{eq:intermediate2} should be used for
various observation times and amplitudes of the background.

\begin{figure}[t]
  \begin{center}
	\includegraphics[scale=0.35]{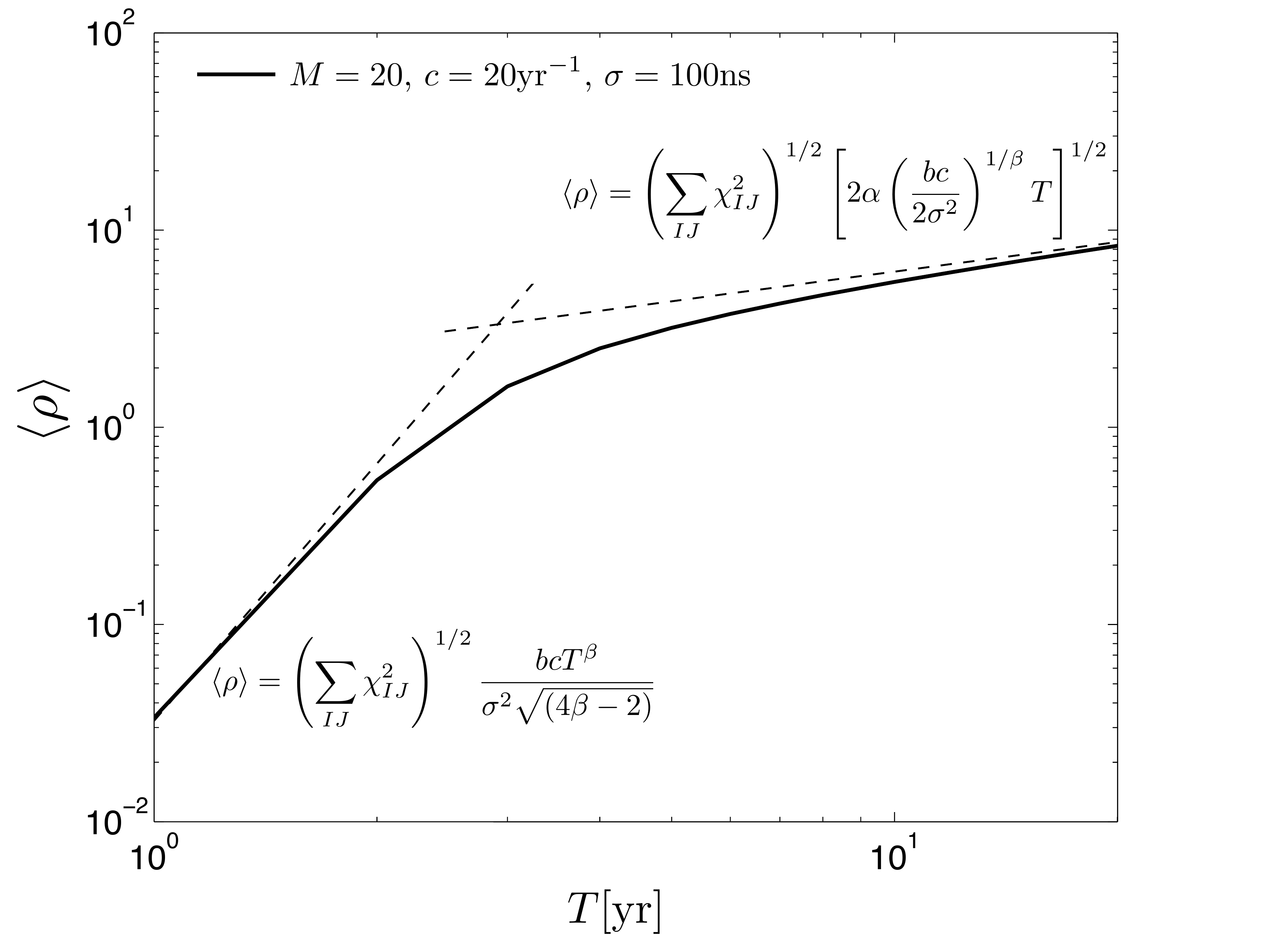}
  \caption{Plot of the SNR versus time in years for
    a stochastic background with amplitude $A=10^{-15}$, $M=20,\,
    c=20~{\rm yr}^{-1}, \, \sigma=100~{\rm ns}$.  The solid curve is
    computed using Eq.~\ref{rhoav} in the time domain using a
    quadratic subtraction timing model. The dashed curves are computed
    using Eqs.~\ref{eq:weaksignal2} and \ref{eq:intermediate2} at late
    times as
    indicated in the plot. Approximating the timing model with a low
    frequency cutoff of $f_L=1/T$ is not quite correct, though it
    does yield the correct dependence on the various PTA
    properties. In this plot we took $T \rightarrow 1.55 T$ for
    both  Eq.~\ref{eq:weaksignal2} and \ref{eq:intermediate2}.}
\label{fig:scalings}
\end{center}
\end{figure}

\begin{figure}[t]
  \begin{center}
	\includegraphics[scale=0.35]{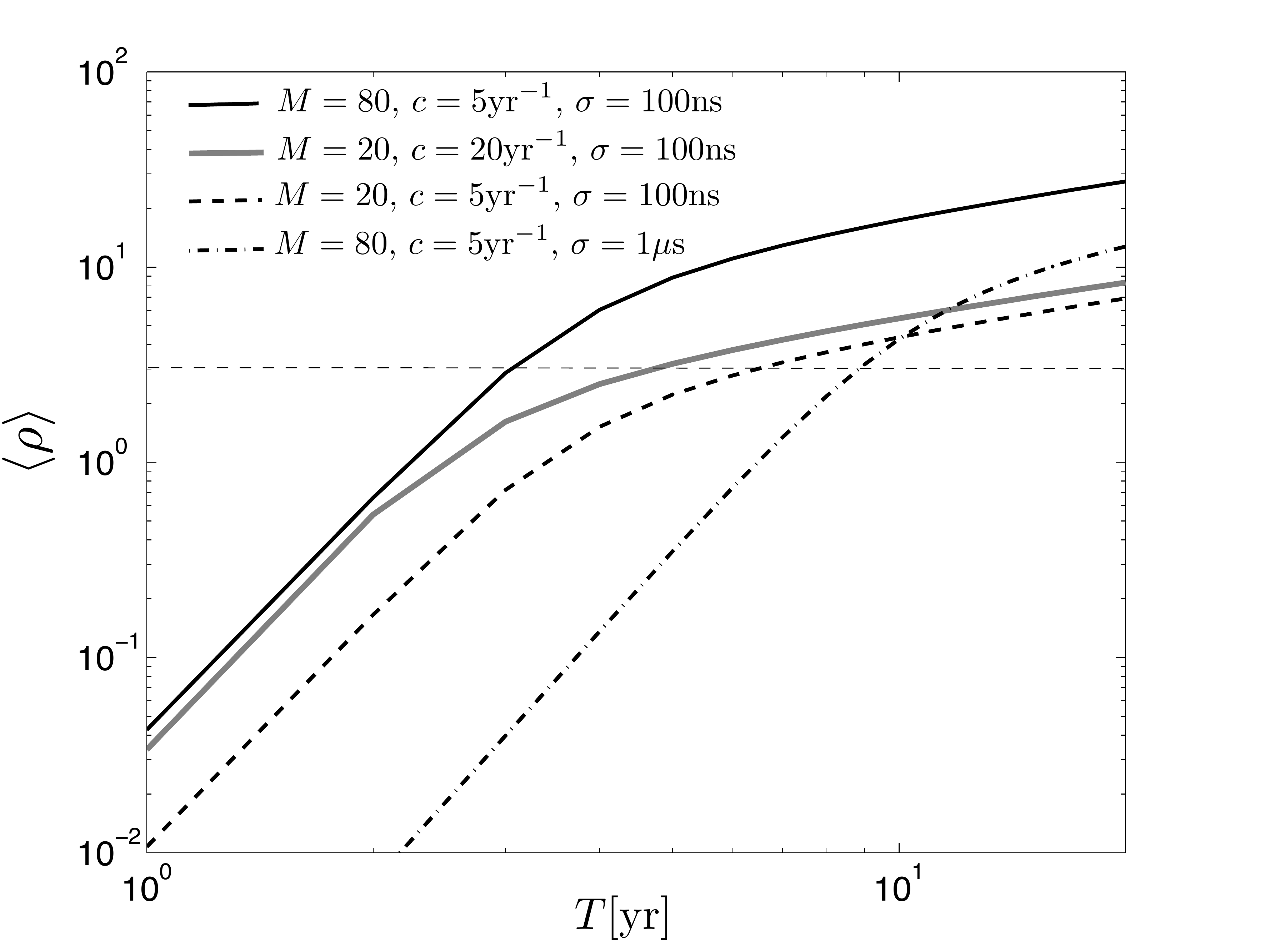}
  \caption{SNR versus time in years for  various
   PTA configurations and an amplitude of the stochastic
    background of $A=10^{-15}$. The solid curve shows a PTA with 80 
    pulsars, timed at a cadence of 5 points per year and an RMS of 
   100~ns. The gray curve shows a PTA with 20 pulsars, an RMS 
   of 100~ns, and a cadence of 20 points per year. The dashed curve
   shows the same PTA with cadence of 5 points per year. Though at early times the
SNR for the gray curve is larger than the dashed curve, at late times
the cadence becomes un-important and the two SNRs
become comparable, as can be seen from Eqs.~\ref{eq:weaksignal2} and
\ref{eq:intermediate2}. The dash-dot curve shows a PTA with 80
pulsars with 1~$\mu$s residuals and a cadence of 5 point per year. Though initially this PTA
has a smaller SNR than the $20$ pulsar PTAs, at late
times it leads to a larger SNR. 
   }
\end{center}
	\label{fig:scenarios}
\end{figure}

Figure~\ref{fig:scalings} shows the average SNR versus time for a GW stochastic
background calculated in the time domain with Eq.~\ref{rhoav}, with  $A=10^{-15}$, $M=20$,
$c=20~{\rm yr}^{-1}$, $\sigma=100~{\rm ns}$ along with the
the two scaling laws Eqs.~\ref{eq:weaksignal2} and
\ref{eq:intermediate2} at late times. For the timing
model we have used quadratic subtraction (fitting for an overall
phase, period, and period derivative). In our frequency domain 
derivation we approximated the
fitting of the timing model with a low frequency cutoff  of $f_L=1/T$.  
In this plot we have taken $T \rightarrow 1.55 T$ for both  
Eq.~\ref{eq:weaksignal2} and \ref{eq:intermediate2} which
reflects the fact that working in the frequency domain with 
a low frequency cutoff at $1/T$ does not
capture the effects of the timing model completely, though it gives the correct
dependence on the PTA parameters.

The difference in the dependence on the observation time, cadence,
white noise RMS of the pulsars, and amplitude of the background of the
two regimes is striking, and critical for PTA optimization.
The combination $bc/\sigma^2$ appears in both cases but with two different powers. While
increasing the cadence and improving the white noise RMS helps greatly
in the weak signal limit, their impact on the SNR in
the intermediate regime is not as significant. The reason for this is
that the SNR is dominated by the lowest frequency
bins. When these bins become gravitational-wave dominated, the
dominant contribution to the noise is the uncorrelated 
pulsar-term GW red noise, rather than the white noise, and this
changes the scaling dramatically: the un-correlated part of the
gravitational-wave stochastic background
is interfering with our ability to detect the correlated piece 
via cross-correlations. The most effective way to beat down the uncorrelated
part of the GW signal is to add more pulsars to the PTA.

PTA data sets are now reaching 20 year time spans with pulsar RMSs at
the level of a few $\mu$s, so the regime where
Eq.~\ref{eq:intermediate2} is valid is probably already relevant or
will soon be. This is critical when considering the kinds of
decisions the pulsar timing community will have to make in terms of 
observing strategies to ensure a prompt and confident detection of the stochastic GW
background.

In Fig.~\ref{fig:scenarios} we illustrate this point. We plot the SNR for a
number of different PTA configurations and a SMBBH stochastic
background with amplitude
$A=10^{-15}$ and slope $\beta=13/3$. The canonical PTA
configuration, with $M =20$ pulsars with $\sigma=100$~ns, and a cadence of $c=20$~yr$^{-1}$
is shown in gray. The dashed curve shows the same
PTA with a cadence of $c=5$~yr$^{-1}$, and though at early times the
SNR for the gray higher-cadence curve is larger than the dashed curve, 
at late times the cadence becomes un-important
as can be gleaned from Eqs.~\ref{eq:weaksignal2} and \ref{eq:intermediate2}.
The solid curve shows a PTA with $M=80$ pulsars timed with a cadence of 5
points per year and an RMS of $\sigma=100$~ns. Both solid curves involve the
same total observation time ($cM=400$). More pulsars and smaller cadence
lead to an SNR that is significantly greater and  a more confident
detection in the long term.  The dash-dot curve shows a PTA with $M=80$
pulsars with $\sigma=1$~$\mu$s residuals and a cadence $c$ of 5 point per year. Though initially this PTA
has a smaller SNR than the $20$ pulsar PTAs, at late
times it leads to a larger SNR because the RMS and cadence become unimportant
compared to the number of pulsars. 

The work we have described in this section began with our attempts to
understand (through simulations) the time it would take a realistic PTA to detect
the gravitational-wave background.  We expected that the weak-signal approximation
would be valid for all times, and in particular that we would benefit
from the dramatic increase in the SNR with time as predicted by 
Eq.~\ref{eq:weaksignal2}. What we found instead was that after some 
time the SNR grows only as the square
root of the observing time. The reason for that, we now understand, is
that the PTA enters the intermediate regime.

\section{Time to detection for realistic pulsar timing arrays}
\label{sec:Timetodetect}

\begin{figure}[t]
  \begin{center}
	\includegraphics[scale=1.05]{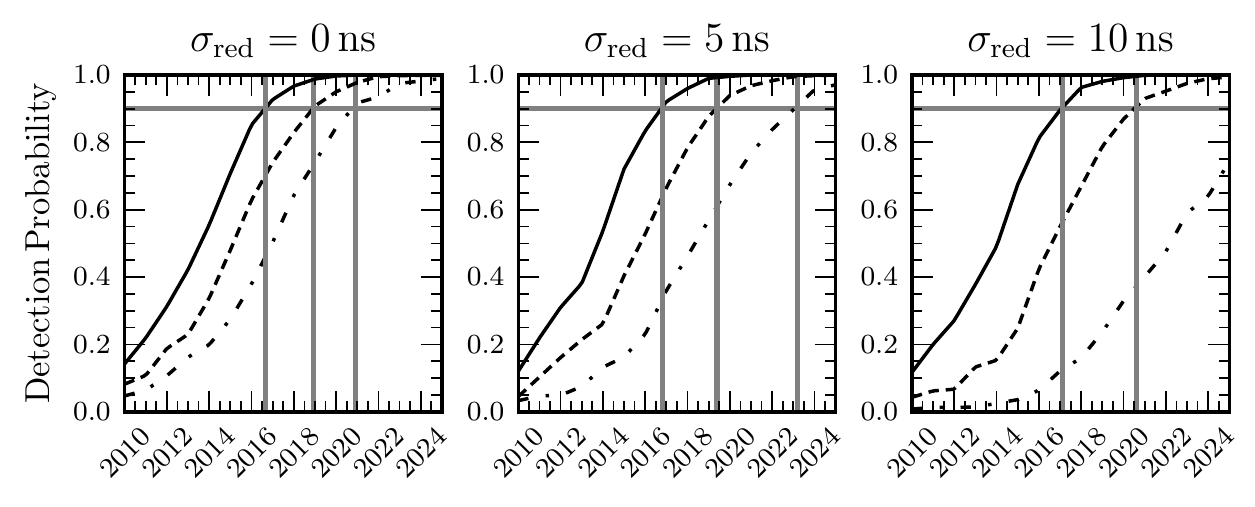}
  \caption{Detection probability versus time in years for
    the NANOGrav pulsar timing array. We show the detection
    probability for 3 different amplitudes $A=5.6 \times 10^{-16}$
    (dash-dot), $A=1 \times 10^{-15}$ (dashed),  and $A=2 \times
    10^{-15}$ (solid) of the stochastic background generated by supermassive
    binary black holes~\cite{s13}.  We have used the combined epoch RMSs of
    the 17 pulsars in~\cite{dfg+12}, started timing 35 total pulsars
    in 2013, and added 3 new pulsars per year with an epoch combined
    RMS of 200~ns prior to hardware improvements 
    (this is the median of the 17 pulsars in~\cite{dfg+12}). 
\label{fig:scenarios1}
  }
\end{center}
	
\end{figure}

In this section we discuss the detection potential of the NANOGrav
pulsar timing array~\cite{jfl+09}.  NANOGrav started collecting data on 17
millisecond pulsars in 2005, and upper limits on
the GW background using these pulsars timed up to 2010 were published
last year~\cite{dfg+12}. Through various surveys, NANOGrav has since increased the 
number of pulsars timed to 35, at about a rate of 3 new pulsars added 
every year. About half of the pulsars are timed at Green Bank, and the
other half at Arecibo.

In our simulations we have used the epoch-averaged combined
residuals for the 17 pulsars in Table 2 of~\cite{dfg+12}.  New
high-bandwidth pulsar timing back-ends have
been installed at the Green Bank Telescope (in 2010), and Arecibo (in
2012).  The new back ends have a bandwith that is about a factor of 10 larger
than the previous ones, and we might expect an increase in TOA
precision of about a factor of 3. In fact, on average only about a
factor of 2 improvement has been observed, presumably due to jitter 
noise~\cite{cs12}.  In our simulations we have assumed a factor of 2
improvement from the new hardware starting in 2010 for pulsars timed
at Green Bank and 2012 for pulsars timed at Arecibo. For new pulsars 
added to the array we have assumed an
epoch combined RMS of 200~ns prior to hardware improvements 
(this is the median of the 17 pulsars in~\cite{dfg+12}).  We have
taken conservative lower ($A=5.6 \times 10^{-16}$),
middle ($A=1 \times 10^{-15}$),  and upper ($A=2 \times 10^{-15}$)  values for the
amplitude of the stochastic background generated by supermassive
binary black holes~\cite{s13}. At each time we perform 1000 injections
of simulated signals. The detection probability is the fraction of injections that
exceed an SNR of 3. The variance of the SNR is quite large in the 
presence of a signal~\cite{ccde+13}, so looking at the SNR threshold alone is not
sufficient, and the 90\% detection probability is typically reached
when the average SNR is significantly larger than 3.

Although most NANOGrav pulsars do not show strong evidence of red
noise, in our simulations we have included the effects of red spin
noise (with spectral index of $-5$)~\cite{cs12}.  We resort to simulations for
this because we have not found a means to include the 
effects of this type of noise into our analytic estimates of the SNR.

Figure~\ref{fig:scenarios1} shows the results of our simulations. The
three panels show red spin noise that induces an RMS of 0~ns, 5~ns,
and 10~ns at 5 yrs. The figure shows that for all but the most
pessimistic scenario (lowest amplitude and largest intrinsic spin noise), a
detection will occur by 2023, and could occur as early as 2016.

\begin{figure}[t]
  \begin{center}
	\includegraphics[scale=0.4]{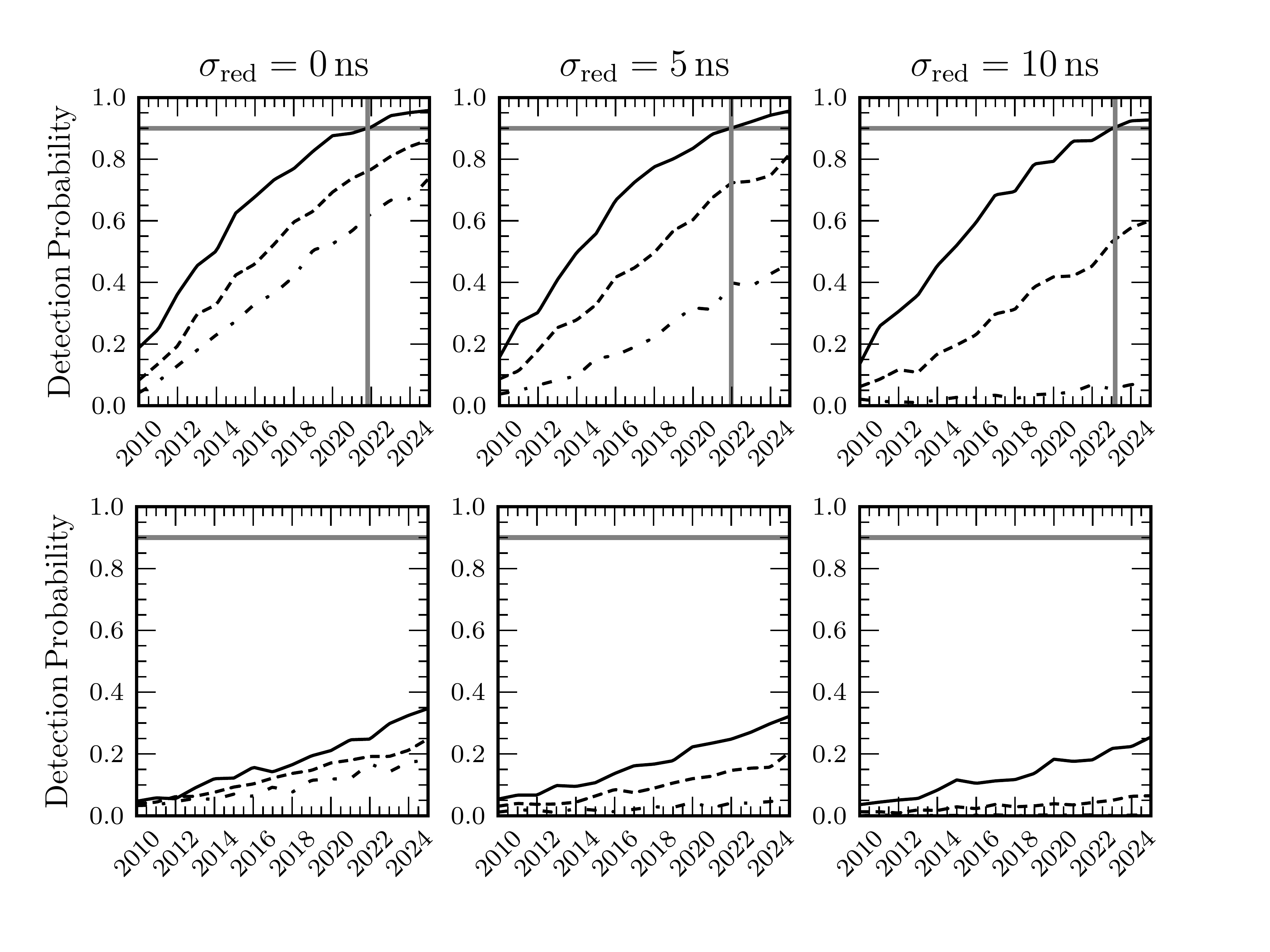}
   \end{center}
  \caption{Detection probability as a function of time for the 17
NANOGrav pulsars of~\cite{dfg+12}, but adding no additional pulsars to
the array (top three panels).
Detection probability as a function of time for 
a pulsar timing array with just 6 pulsars with white
noise RMSs of 10~ns (bottom three panels). 
The detection
    probability is shown for 3 different amplitudes $A=5.6 \times 10^{-16}$
    (dash-dot), $A=1 \times 10^{-15}$ (dashed),  and $A=2 \times
    10^{-15}$ (solid) of the stochastic background generated by supermassive
    binary black holes~\cite{s13}.}
  \label{fig:timetodetection}
\end{figure}

This prompt detection is possible because of the addition of new
pulsars to the timing array. The importance of adding new pulsars is
illustrated in Fig.~\ref{fig:timetodetection}. The top three panels
show the detection probability as a function of time for the 17
pulsars in~\cite{dfg+12}, but adding no additional pulsars. The bottom
three panels show a pulsar timing array with just 6 pulsars, with white
noise RMSs of 10~ns.  Clearly, increasing the number of pulsars is
critical to our ability to confidently detect the stochastic background.

\section{Summary}
\label{sec:conc}

We have produced analytic expressions for the SNR of the optimal
cross-correlation statistic for stochastic background searches
as a function of the various PTA properties. We find that the weak
signal limit scalings where $\langle \rho \rangle \propto T^\beta$ do
not apply at all times.  Once enough time has elapsed the lowest
frequencies of pulsar timing residuals become GW-dominated and the
scaling changes to $\langle \rho \rangle \propto \sqrt{T}$.  In
addition, the dependence on the cadence and the RMS of the residuals
weakens significantly, and the best strategy to confidently detect
the stochastic background is to increase the number of pulsars in the PTA.

PTAs may enter this new regime more quickly than one might have thought
naively. Of the 17 pulsars in the 5 year data set analyzed
in~\cite{dfg+12}, 9 are already in this regime for typical values of the amplitude of the
stochastic background ($A \sim 10^{-15}$). Some current PTA data sets span close 
to 20~yrs, and for the same amplitude of the background their lowest frequency bins
are GW-dominated if the RMSs of the timing residuals are smaller than 
a few $\mu$s.

We have also made realistic simulations of the NANOGrav PTA. 
NANOGrav is currently timing 35 pulsars and adding more to its array at a rate
of about 3 per year. Where possible, we have
used the measured RMSs of millisecond pulsars, and for future
discoveries we have used the median RMS of the existing set. Most
NANOGrav pulsars do not show strong evidence for intrinsic red 
noise. Nevertheless, to be conservative, we have included the effects of red spin noise in
our simulations. We find that a confident detection of the background could occur as early as 2016.

It is worth pointing out that although for stochastic background searches the RMS of timing
residuals becomes un-important at late times, there are very good 
reasons to improve on the RMS of pulsars. For example, for other types of GW signals such as 
continuous waves and bursts, the sensitivity is dominated by 
the pulsars with the smallest residuals. 

\ack
We would like to thank Jim Cordes, Paul Demorest, Scott Ransom, Rutger van Haasteren, and the
members of the NANOGrav detection working  group for
their comments and support. The work of XS and JE was partially funded
by the NSF through CAREER award number 0955929, PIRE award number 0968126, 
award number 0970074, and the Wisconsin Space Grant Consortium.
JDR would like to acknowledge NSF award CREST-0734800.

\section*{References}
\bibliographystyle{iopart-num} 
\bibliography{apjjabb,bib}

\end{document}